# p=constant compression on loose Hostun sand:

# The case of an anisotropic response

## P. Evesque

Lab MSSMat, UMR 8579 CNRS, Ecole Centrale Paris
92295 CHATENAY-MALABRY, France, e-mail evesque@mssmat.ecp.fr

**Abstract:**

*Experimental data from axially symmetric compression test at constant mean pressure $p=(\sigma_1+\sigma_2+\sigma_3)/3$ on Hostun sand from Flavigny experiments on loose sands are used to study the validity of an "isotropic" modelling at different densities . It is found that the material response is not isotropic even at small deviatoric stress. As an "isotropic" behaviour is found for compression test at constant volume on the same sand, this new result questions the unicity of the trajectory in the classical phase space of soil mechanics (q,p,v) ,with $q=\sigma_1-\sigma_2$, v=specific volume, and asks whether this space shall be taken larger than 3d or not.*

**Pacs # :** 5.40 ; 45.70 ; 62.20 ; 83.70.Fn

______________________________________________________________________

Recent papers [1-4] have proposed a new simple description of the behaviour of granular media based on an "isotropic"[5] incremental modelling. More recently, experiments on clay [6,7] and on Hostun sand [8,9] have been used to check the validity of this modelling. These works have concluded that the "isotropic" modelling is valid when the deviatoric stress $q=\sigma_1-\sigma_2$ remains small, *i.e.* $q<\sigma_2=\sigma_3$ (or q<0.5p).

In this paper, the study is pursued; and we examine other experimental results of compression test at constant mean pressure $p=(\sigma_1+\sigma_2+\sigma_3)/3$ on Hostun sand. These ones have been performed by an other experimentalist [10]. It is found that these data do not agree with the "isotropic" modelling even at the beginning of the compression when the deviatoric stress $q=\sigma_1-\sigma_2$ remains very small. Few explanations are given, based either on a non isotropic initial assembly, or on a spontaneous symmetry breaking of the mechanical response, or on the need to introduce a phase space with a larger dimension than the usual one, *i.e.* (q,p,v).

Furthermore, results from the same experimentalist, *i.e.* E. Flavigny, on compression test on Hostun sand at constant volume are briefly recalled. It is shown that some of these results do agree with an "isotropic" modelling; but some others do not. This means that these samples, which are supposed to be rather identical, do exhibit at least two different kinds of behaviour so that they shall lead to the conclusion that (i) the trajectory in the (q,p,v) space is not unique for a given set of initial condition $(q_o=0,p_o,v_o)$ and consequently (ii) that the classical phase space (q,p,v) is not large enough to include these effects so that the real phase space shall have a dimension larger than 3. The paper is built as follows: we first recalled the basis of the modelling; then we describe the experimental result; we discuss them and conclude.

In the general frame work of incremental modelling [1-4] with a small number of zones, one expects that all compression tests performed on a given sample obeys an equation of the kind of Eq. (1), where all coefficients $C_o$, $\alpha$, $\nu$, $\nu'$ and $\nu''$ varies as the compression proceeds. At this stage, it is worth recalling that $\nu$, $\nu'$ and $\nu''$ shall be different in the most general case, but that $\nu=\nu'$ if the two different compression paths ($\delta\sigma_1 \neq 0$, $\delta\sigma_2=0$, $\delta\sigma_3=0$) and ($\delta\sigma_1=0$, $\delta\sigma_2 \neq 0$, $\delta\sigma_3=\delta\sigma_2$) pertain to the same zone. It is also worth recalling that any axially symmetric compression gives access only to the quantity $\alpha-\nu''$ ; so $\alpha$ and $\nu''$ cannot be measured separately with such axi-symmetric triaxial apparatus, so that one can choose $\alpha=1$ when one limits the experimental range to axial symmetry.

$$\begin{pmatrix} de_1 \\ de_2 \\ de_3 \end{pmatrix} = -C_o \begin{pmatrix} 1 & -n' & -n' \\ -n & a & -n'' \\ -n & -n'' & a \end{pmatrix} \begin{pmatrix} ds_1 \\ ds_2 \\ ds_3 \end{pmatrix} \qquad (1)$$

Furthermore, one can ask whether it is useful to use Equation (1) rather than its simplified "isotropic" version which states $\nu=\nu'=\nu''$ and $\alpha=1$. Indeed, it has been argued in [6-8] that the validity of the "isotropic" modelling can be checked by applying an axially symmetric triaxial compression at constant pressure $p=(\sigma_1+\sigma_2+\sigma_3)/3$, since one expects that $\delta v=0$ when $\delta p=0$ in the case of an "isotropic" response whatever the applied deviatoric stress $q=\sigma_1-\sigma_2$.

Fig. 1 reports the variations of $q/\sigma_3$, of volume deformation $\varepsilon_v$ and of $\partial\varepsilon_v/\partial\varepsilon_1$ as a function of the axial deformation $\varepsilon_1$ for different compression tests at constant mean pressure p on the same Hostun sand packed at different initial specific volume v (or void index e such as $v=1+e$) and under different mean stress. The range of investigated pressure is 0.7 atmosphere to 6 atmospheres and of initial specific volume v is 1.92 to 2, which corresponds to a range of the porosity $\Phi=e/(1+e)$ from 0.479 to 0.502. This range corresponds to the so-called loose sand regime. Conventions are as follows: positive volume- and positive axial- deformations correspond to volume- and height- decreases.

Data show :
- that the sample contracts always, whatever the mean pressure p and the initial density, in the investigated range, (this one corresponds to loose sample)
- that the maximum of volume $\partial\varepsilon_v/\partial\varepsilon_1$ variation occurs at small deformation $\varepsilon_1$, when q=0
- that the volume variation decreases $\partial\varepsilon_v/\partial\varepsilon_1$ continuously to 0. Its typical maximum value ranges in between $\partial\varepsilon_v/\partial\varepsilon_1=[0.3 , 0.9]$ depending on the test.

Considering the experimental condition $\delta p=0$, which imposes $\delta\sigma_1=-2\delta\sigma_2$, Eq. (1) leads to $\partial\varepsilon_v/\partial\varepsilon_1=(1-\alpha-2\nu+\nu'+\nu'')/(1-\nu)$. Assuming then $\nu=\nu'$, this can be rewritten as $\partial\varepsilon_v/\partial\varepsilon_1 =1-(\alpha-\nu'')/(1-\nu)$. As experiments show that $\partial\varepsilon_v/\partial\varepsilon_1=[0.3 , 0.9]$, it means that $\alpha-\nu''$ shall be much smaller than $1-\nu$ at small deformation.

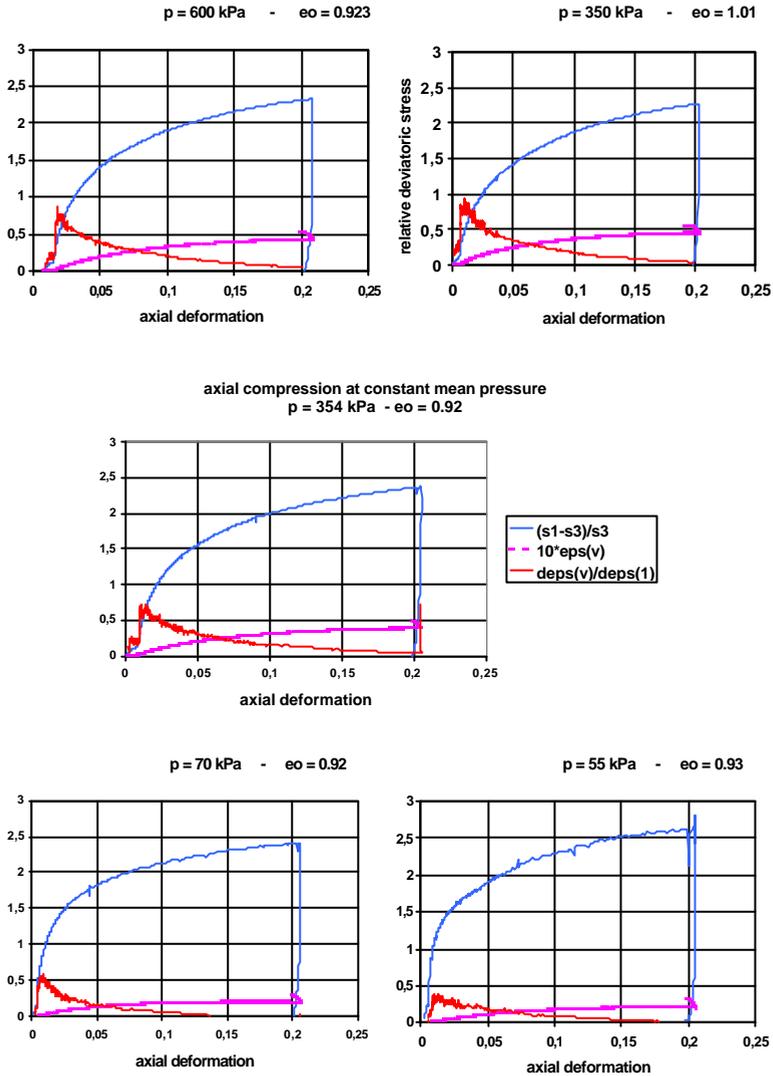

**Figure 1:** *variations of the deviatoric stress* $q/\sigma_3$ *(blue), of* $\partial\varepsilon_v/\partial\varepsilon_1$ *(red) and of volume deformation* $\varepsilon_v$ *(pink) vs. the axial deformation* $\varepsilon_1$ *during axial compressions at constant mean pressure* p*, for different* p *values and for different initial densities, on Hostun sand. Data from* [10]*; the void index* e *is related the specific volume* v *by* v=1+e *and to the porosity* $\Phi$ *by:* e=$\Phi$/(1-$\Phi$).

**up-left Fig.:** (p=600 kPa, $e_o$=0.923); **up-riht:** (p=350 kPa, $e_o$=1.01); **centre:** (p=354 kPa, $e_o$=0.92); **bottom-left:** (70 kPa, $e_o$= 0.92); **bottom-right:** (p=55 kPa, $e_o$=0.93) .

*Positive volume- and axial- deformations correspond to volume- and height- decreases.*

So these results demonstrate that the response never obeys Eq. (1) with the "isotropic" assumption, even at small deformation. Indeed, the fact that $\partial\varepsilon_v/\partial\varepsilon_1$ tends to 0 for large axial deformation $\varepsilon_1$ does not imply an "isotropic" response; it is simply linked to the existence of the "critical" state and to the fact that the sample tends to a typical volume which does not evolve with deformation; in other words, it means simply that the typical specific volume is reached within a finite deformation; one way to describe this result is just to assume $\nu=1/2$ when the material response can be described by the "isotropic" response, *i.e.* when $\alpha=1$, $\nu=\nu'=\nu"$ , or more generally by $(-\alpha+2\nu-\nu'-\nu")=0$, when the response is not "isotropic".

So the above results of Fig. 1 are in complete contradiction with previously reported ones [6-9] on clays [6,7] and on Hostun sand [8,9] since they demonstrate that the response is not "isotropic" . These results are rather surprising since the material is supposed to be built using a process which tries keeping the material as homogeneous and as isotropic as possible, and since the initial stress state is isotropic. Indeed, this should imply that the material response shall be isotropic in the vicinity of $\sigma_1=\sigma_2=\sigma_3$.

In previous papers [1-3], it was also argued that an anisotropic response should be induced by a change of the contact distribution, so that it could happen only after some large axial deformation. However, our study in [6-8] has concluded that this assumption is wrong, since experiments in [6-8] are characterised by an anisotropic response as soon as q/p>0.5 *even* if the axial deformation is remaining quite small.

One can try to explain the results of Fig. 1 within different schemes:
(1) The simplest way to explain these results is to assume the existence of some initial anisotropy; this one can be generated by the generation of a non random distribution of contact orientations during the building process of the pile.
(2) An other possible explanation is to assume a spontaneous breaking of symmetry of the response as soon as q/p is non zero; in this case the axis 1 will be differentiate from the axes 2 and 3 by the existence of a quite small but finite q/p. Even if this mechanism looks rather surprising in the incremental approach, one shall note that the concept of perfect plasticity with a single mechanism allows such a hypothesis: at some stage of the deformation process, the plastic mechanism is activated, which imposes a flow rule different from what was observed before this activation. In the present case, it would mean that the activation threshold is at q/p≈0.
(3) One can perhaps attribute the anisotropic response of Fig. 1 as linked to the fact that the trajectory crosses the limit between two different zones characterised by two different incremental responses so that one could have $\alpha=1$, $\nu=\nu"$, but $\nu\neq\nu'$ , which leads to $\partial\varepsilon_v/\partial\varepsilon_1=(1-\alpha-2\nu+\nu'+\nu")/(1-\nu)=(\nu'-\nu)/(1-\nu)$. So, this could explain Fig. 1.

**Variability of the mechanical behaviour:** It is worth noting that undrained compressions performed by the same author on Hostun sand, or by other authors on the same sand or on other sands, provide sometimes a trajectory which starts at constant mean pressure p, but sometimes the mean pressure decreases at once on the contrary. Both behaviours are observed for very loose or loose samples more often than for denser ones, *cf.* p. 72-73 & 176-183 of ref. [11]. Sometimes also, in the case of dense samples, one can observe an initial increase of the mean pressure during the first stages of an undrained compression (it is assumed in this case the presence of air in the pores increases the "effective" compressibility of the effective fluid which fills the pores; this finite compressibility makes the test not capable to measure the effective change of volume of the pore space, so that this test looks like a test at constant water pressure, *i.e.* at $\sigma_2 = \sigma_3 = c^{ste}$, rather than at undrained condition).

Anyhow, results from undrained compressions can and shall be interpreted in the scheme given by Eq. (1) too. So they shall be interpreted within the same scheme as the compression at constant mean pressure. This is done below:

One remarks then that all these compressions are performed with the same constrain of axial symmetry, *i.e.* $\sigma_2 = \sigma_3$ ; so, the number of real input parameters is 2, *i.e.* (q & p), or (q & $\varepsilon_v$), or ($\varepsilon_1$ & $\varepsilon_v$) ,…; it means in turn that the evolution of a given sample is described by a 2d surface. One shall also take into account the fact that the classical phase space of soil mechanics , *i.e.* the space of minimum dimension which is able to provide a complete representation of the sample evolution, is the 3d space defined by (q,p,v).

It results from the analysis of the previous paragraph that the sample evolution is completely determined by the knowledge of $\delta q$, $\delta p$; but the way to get the same evolution can be multiple, either by controlling directly ($\delta q, \delta p$) or by controlling any other couple of parameters ($\delta q, \delta v$) which imposes the same ($\delta q, \delta p$). This results from the hypothesis of a determinist evolution. It means in particular that if one observes some $\delta v_o$ evolution during a compression test at ($\delta q_o, \delta p=0$), one shall observe dp=0 during a compression which imposes the incremental path ($\delta q_o, \delta v_o$) to the same sample. This imposes in turn that classical undrained compression which results in the path ($\delta q_o, \delta v_o=0, \delta p=0$) shall be also obtained via a compression at constant mean pressure which imposes ($\delta q_o, \delta p=0$). In the same way and more generally, one shall conclude that an undrained compression test which results in the following path ($\delta q_o, \delta v_o=0, \delta p<0$) shall lead to a compression path at constant pressure characterised by ($\delta q_o, \delta p =0, \delta v_o<0$) in virtue of the action-reaction principle.

So, in conclusion, if one follows these guide lines, one shall associate the anisotropic response exhibited by Fig. 1 of the present paper to the anomalous anisotropic behaviours observed some times in loose sand during undrained compression.

However, it is worth noting that the existence of an initial anisotropy is not the only scheme of interpretation. For instance, these experiments could question the validity of the Terzaghi principle (this principle states that neither the water pressure $u_w$ nor its variation $\delta u_w$ is involved directly in the mechanics of granular matter, so that the only parameters which control the granular-medium deformation are the three effective stresses $\sigma_1, \sigma_2, \sigma_3$, and their increments $\delta\sigma_1, \delta\sigma_2, \delta\sigma_3$). If this principle was not holding true, the number of independent parameters would be 4 instead of 3, which will be noted ($\delta\sigma'_1, \delta\sigma'_2, \delta\sigma'_3, \delta\sigma'_4$); (in the case when the Terzaghi hypothesis is satisfied, these 4 parameters lead to ($\delta\sigma_1, \delta\sigma_2, \delta\sigma_3, \delta u_w$) for which $\delta u_w$ does not act on the granular medium deformation). Moreover, one shall take into account the condition of axial symmetry which imposes some relation between the 4 parameters ($\delta\sigma'_1, \delta\sigma'_2, \delta\sigma'_3, \delta\sigma'_4$); this reduces to 3 the number of independent parameters which control any real trajectory corresponding to axially symmetric deformation, if the Terzaghi principle was not satisfied. This 3d space shall be a sub-space of the real phase space; as this last one shall contain also the specific volume, the real phase space should be at least 4d.

Indeed this approach and the other ones, labelled from 1 to 3 in the above paragraphs, can be described within a single scheme, just by assuming that the mechanical response is sensitive to other parameters. In this case, one shall add to the classical phase space, defined by (q,p,v), few other dimensions which describe the evolution of the other independent control parameters. For instance, if the anisotropic response is due to a pre-existing anisotropy of the assembly, the 4$^{th}$ dimension would be just the parameter which defines the anisotropy of the assembly and its evolution.

At last, it is possible that , as the compression proceeds, the sample becomes inhomogeneous and/or the stress field generates non diagonal stress term locally. Such effects are known already to appear; they can lead to strain localisation, for instance. Indeed generation of inhomogeneities can break the initial axial symmetry spontaneously rather rapidly. Such inhomogeneities can sometimes be taken into account by the variation of the local specific volume v; however it can require also to introduce a new local parameter. Anyhow, it requires surely to introduce (i) the 3d real space (x,y,z), (ii) the local evolution of the (q,p,v) sub-set in this (x,y,z) space and (iii) the space coupling between the (q,p,v) subsets which allow to describe the complete deterministic evolution after a local perturbation. Indeed evolution due to spatial coupling can generate a change of (i) the principal directions of the local stress tensor, (ii) of the local contact distribution and (iii) of the extra parameters needed to describe localisation; so the situation appears rather complex and generation of inhomogeneities requires to increase undoubtedly the dimension of the real phase space.

As a conclusion, this paper has studied the response of loose-sand samples to compressions at constant mean pressure p in the case when the mechanical response exhibits a volume decrease at the early stage of the deformation. Parallel with anomalous behaviours observed on undrained compressions has been done and the source of this anomalous behaviour has been attributed to the anisotropy of the initial assembly. However, other possible explanations have been listed, they assume in general the need to increase the dimension of the real phase space which is able effectively to describe the mechanics of granular matter . At last it is concluded that more experimental results exhibiting such anomalous behaviours are required, and should be investigated, prior to conclude definitively.

***Acknowledgements:*** CNES is thanked for partial funding. Professor Etienne Flavigny is kindly thanked for having provided his experimental data, and for stimulating discussions.

The electronic arXiv.org version of this paper has been settled during a stay at the Kavli Institute of Theoretical Physics of the University of California at Santa Barbara (KITP-UCSB), in june 2005, supported in part by the National Science Fundation under Grant n° PHY99-07949.

*Poudres & Grains* can be found at :
http://www.mssmat.ecp.fr/rubrique.php3?id_rubrique=402